\documentclass{ptapap}

\author{{\L}ukasz Wyrzykowski}[OAUW]
\affil[OAUW]{Warsaw University Astronomical Observatory, Al. Ujazdowskie 4, 00-478 Warszawa, Poland}

\title{First year of the Gaia Science Alerts}

\begin{document}

\maketitle

\begin{abstract}

Since mid 2014 Gaia mission delivers daily millions of observations of the whole sky. Among them we search for transient events, e.g., supernovae, microlensing events, cataclysmic variables, etc. 
In my talk I describe the near-real-time Gaia data processing pipeline for anomaly detection, show first scientific results from years 2014/2015 and describe the organization of the ground-based network for the photometric and spectroscopic follow-up of Gaia alerts.

\end{abstract}

\section{Introduction}

Gaia is the premier ESA's mission, launched in December 2013. Its primary goal is to observe more than one billion stars in the Milky Way and provide superb astrometry, photometry and astrophysical parameters for each one of them. Because the entire sky is being scanned multiple times (40 to 200 times, being 70 the average), Gaia is also a time-domain all-sky survey. Thanks to efficient on-board object detection system, Gaia is capable of autonomously detecting newly appearing objects, as contrary to its predecessor, the Hipparcos mission from 1990s. This feature of Gaia enables it to discover transient objects, which appear temporarily on the sky. The Gaia Science Alerts group, based in Cambridge (UK) and Warsaw (Poland), is given a task of rapid analysis of daily Gaia data deliveries. From an average of 50 million observations per day, the group needs to identify and characterize the anomalous or new astrophysical sources. These objects require imminent additional follow-up observations, as their scientific potential degrades very quickly and may be lost forever. This includes, for example, supernovae (SNe), cataclysmic variables (CVs), tidal disruption events (TDE), microlensing events and other rare and unusual astrophysical events. Over its 5 years mission, Gaia is expected to detect about 6000 supernovae of all types down to 19 mag, few thousands of CVs, a hundred TDEs and about a 1000 of photometric microlensing events.
Here we present a brief summary of the operation of the first year of the Gaia Alerts and its results. 

\section{Gaia Alerts detection system}

Gaia's on-board instrument is a 1 Gigapixel camera, containing 106 CCDs. Due to constrains on the bandwidth and the fact that most of the sky is empty (for the satellite $\sim$20.5 limiting magnitude), Gaia does not transmit pixels, but just small cutouts (windows) around each detected source. During each scan, as the source crosses the focal plane (Gaia operates in a drift-scan mode), the position and brightness of each source is measured on 9 Astrometric Field (AF) CCDs, followed by low-dispersion blue and red spectro-photometers (called BP/RP) and, for stars brighter than 16 magnitudes, also by a high-dispersion radial velocity spectrographs (RVS). It takes about 45 seconds for a source to cross the focal plane. Because of the spinning of the satellite, the observations of the same source are likely to come from the second telescope after 106 minutes (unless the spin axis precesses away from that source). Another visit of the same part of the sky typically happens after 30 days, on average. Such sampling pattern is optimal for discovery, however additional follow-up observations are necessary in order to provide more detailed information and classification for each source.

\subsection{Detection}
The AlertPipe software is run in the Institute of Astronomy, Cambridge, each time there is a new delivery from the first data reduction, provided by the Initial Data Treatment (IDT) at ESAC (Spain). The data is being ingested into the database and the detection system is run. New transient detections may appear from two different sources: new source or old source alert. A new source alert is generated when a new object appears at a position with several previous non-detections. Old source alerts are raised when a variation on the source magnitude is perceived. Different criteria are applied in order to pre-select candidates for alerts. Figure 1 shows a schema of the detection process within AlertPipe.

\begin{figure}
\includegraphics[width=\textwidth]{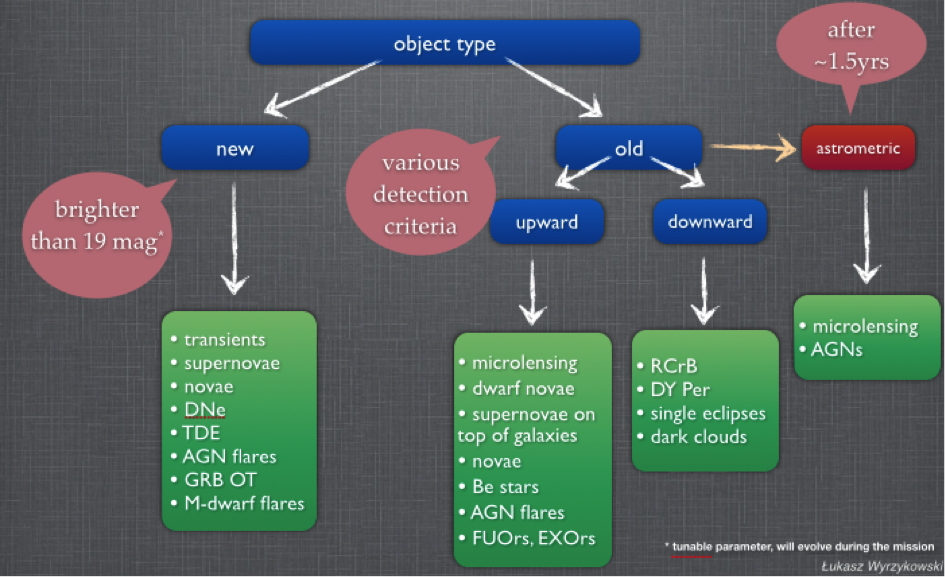}
\caption{Gaia Science Alerts detection pipeline schema}
\label{fig:detection}
\end{figure}

Each candidate is then analyzed internally, by using available Gaia data. If the object is new, has at least two detections and has no flags for being an artefact, it is labeled as TRANSIENT. If the object already has an observing history, but there are two consequent outlying observations, it is labeled as BUMP or DIP, depending on the direction of the anomaly (got brighter or fainter, respectively). 

\subsection{Classification}

The contextual data can provide important information about the nature of an object. Each candidate surviving the detection step is then cross-matched with locally available archival databases. Those include SDSS, DSS2, GSC2, USNOB, Leda, Veron, OGLE and APASS among others. For example, bright stars are being identified in vicinity of a candidate alert and their position is used to rule out artefacts due to bright stars spikes. If a galaxy is present nearby to the position of the Gaia alert, then the candidate is classified as an extragalactic transient, most likely a supernova. If the archival image has no objects at a given position, an orphan transient is found, often caused by a supernova in a low luminous host or a cataclysmic variable. Cross-match is also performed against catalogues of known variable stars, e.g. from CRTS, ASAS or OGLE surveys. If an alert matches the position of a known variable star, it is further ignored and no alert is raised.

\subsection{Spectral classification with BP/RP}
An additional powerful feature of Gaia as a transient survey is the presence of low-resolution (R$\sim$100) BP/RP data. As shown in \cite{Blagorodnova}, Gaia's BP/RP spectra can be used for unambiguous classification of transients into major classes, e.g. SNe, stars, AGN and CVs. Moreover, for objects brighter than $\sim$18.5 mag, the spectra can also be used for more detailed analysis of supernovae into types Ia/II and provide estimation on their redshift and epoch to peak brightness. 

\section{First Year of Alerts}
Gaia Science Alerts have started its routine operation in July 2014 and the alerts has been reported on the web page: http://gaia.ac.uk/selected-gaia-science-alerts.
Figure 2 shows the sky coverage obtained by Gaia in January 2016. By this date, the entire sky was observed at least once and some parts even significantly more often. This allowed for unambiguous detection of new sources using the information on the history of previous non-detections. The first confirmed supernova found by Gaia was Gaia14aaa (Fig. 3), discovered on 30 August 2014 in a region with relatively high cadence of observations.

Since then, until mid-2015 Gaia has detected 274 transients. Vast majority of them were supernovae, primarily of Type Ia, however, there were also detections of cataclysmic variables and candidates for microlensing events. 
In September 2014 Gaia has detected a very rare type of cataclysmic variable of AM CVn-type (See Fig. 4 and \citealt{Campbell}).

\begin{figure}[h]
\centering
\includegraphics[width=\textwidth]{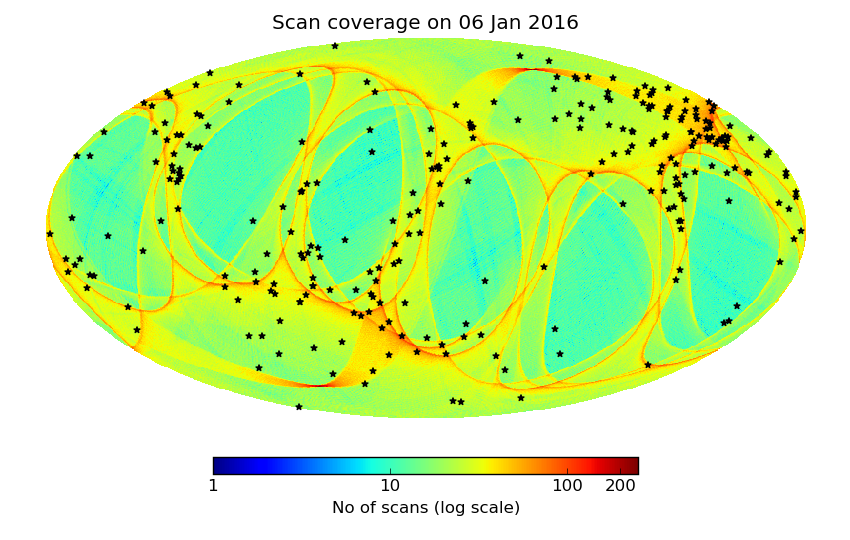}
\caption{Map of the sky coverage by Gaia in January 2016, colours indicating the number of observations collected so far. The stars mark the positions of all 274 alerts found during the first year of operation of the Gaia Science Alerts. From http://gsaweb.ast.cam.ac.uk/alerts/maps/alerts-equatorial-coverage-map.png.}
\label{fig:map}
\end{figure}

\begin{figure}[h]
\includegraphics[width=\textwidth]{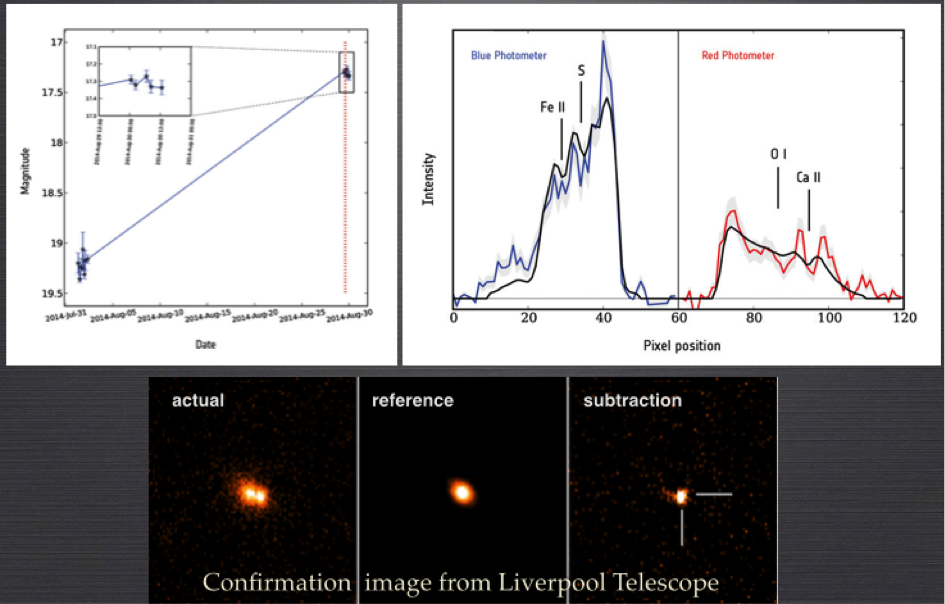}
\caption{First Gaia's supernova, Gaia14aaa. Upper left figure shows the light curve, where historic observations of the galaxy are at a level of 19 mag and the red vertical line shows the alert trigger at 17.3 mag. The BP/RP spectrum (upper right) indicated a good match to type Ia supernova (model shown by the black line). The confirmation image with the Liverpool Telescope was obtained a couple of days after the detection by Gaia (bottom panel).}
\label{fig:first}
\end{figure}

\begin{figure}[h]
\includegraphics[width=\textwidth]{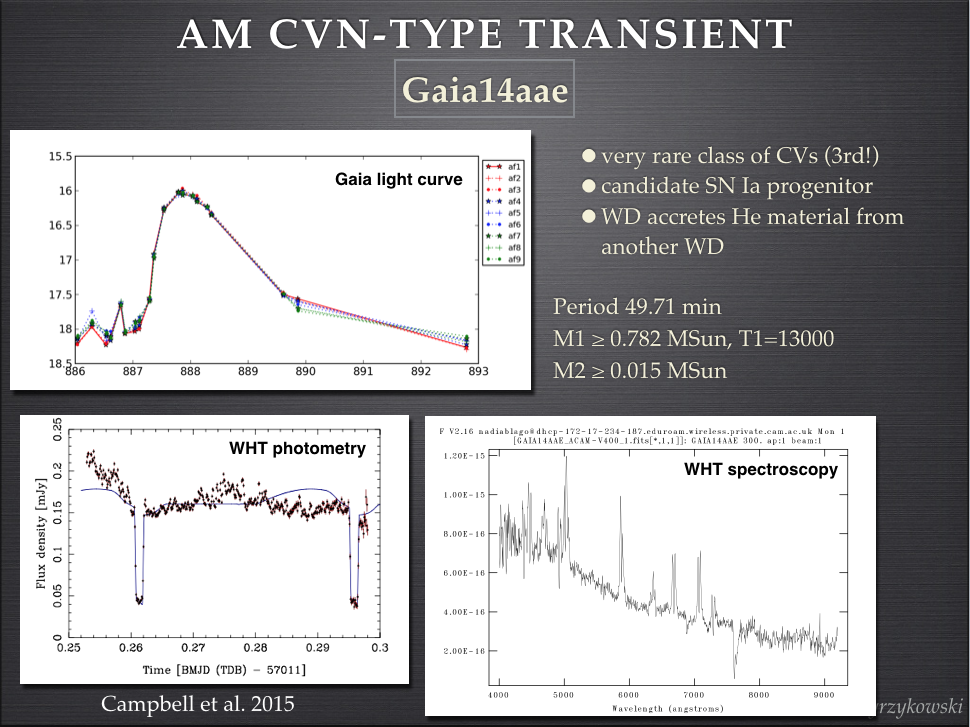}
\caption{Gaia14aae, AM CVn-type transient, found by Gaia and ASAS-SN projects.}
\label{fig:gaia14aae}
\end{figure}

\section{Ground-based follow-up}
Apart from spectroscopic follow-up carried out on multiple telescopes around the world, the long-term multi-band photometric follow-up has been carried out by the network supported by the OPTICON FP7 EC's grant. The network consists of about 15 active telescopes ranging in size from 0.5m to 2m\footnote{List of active partners: https://www.ast.cam.ac.uk/ioa/wikis/gsawgwiki/index.php/Verification\_phase}. The photometric data is reduced and calibrated in an automated fashion using Cambridge Photometric Calibration Server\footnote{http://gsaweb.ast.cam.ac.uk/followup/} designed by Sergey Koposov and Lukasz Wyrzykowski. The server yields a roughly calibrated homogenous data from multiple telescopes and multiple observational configurations. An example of a follow-up light curve is shown in Fig. 5.

\begin{figure}[h]
\centering
\includegraphics[width=7cm]{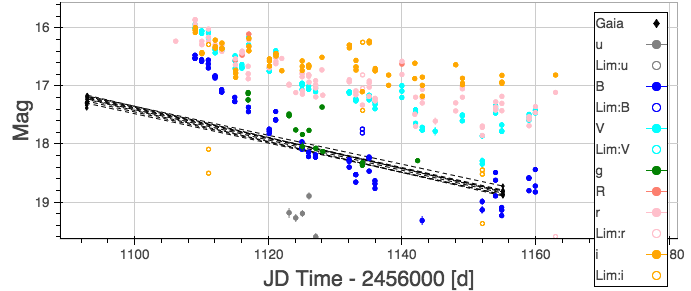}
\caption{Example of multii-site, multi-band photometric follow-up of Gaia15adb, SN Type Ia, carried out by the OPTICON-supported network of ground-based telescopes. }
\label{fig:follow-up}
\end{figure}

\section{Conclusions}

With the Gaia Science Alerts, Gaia has already started delivering interesting scientific results. The number of transients detected in the following years is expected to grow significantly, thanks to longer baseline of observations and continuous improvements being made to the detection pipeline and better data understanding. We would like to encourage everyone interested in Gaia transients to join the Gaia Science Alerts community in order to fully exploit the potential of this mission. More details, as well as archives of presentations from Gaia Science Alerts Workshops, held since 2010, are available on the website: 
http://www.ast.cam.ac.uk/ioa/wikis/gsawgwiki.

\acknowledgements{This talk relies on work of numerous people. First I acknowledge the entire Data Processing and Analysis Consortium of Gaia (DPAC), and in particular Cambridge Gaia Science Alerts Group including: Simon Hodgkin, Guy Rixon, Diana Harrison, Arancha Delgado, Heather Campbell, Sergey Koposov, Nadia Blagorodnova, Morgan Fraser, Goska van Leeuven and many others. From Warsaw I would like to thank Zuzanna Kostrzewa-Rutkowska, Krzysztof Rybicki, Marzena Sniegowska, Krzysztof Ulaczyk, Michal Pawlak, Krystian Ilkiewicz, Aleksandra Hamanowicz, Jakub Klencki, Piotr Wielgorski and many others. We thank the entire Gaia Alerts follow-up network. 
This work has been supported by Polish NCN 'Harmonia' grant No. 2012/06/M/ST9/00172 as well as OPTICON FP7 EC grant no. 312430 and Polish MNiSW W32/7.PR/2014. 
}

\bibliographystyle{ptapap}
\bibliography{wyrzykowski-gaia}

\end{document}